\documentclass[a4paper]{article}

\usepackage{INTERSPEECH2022}
\usepackage{amssymb,amsmath,epsfig}
\usepackage{textcomp}
\usepackage{amsfonts,amsthm,amsopn}
\usepackage{url}
\usepackage[capitalise]{cleveref}
\usepackage{graphicx,caption,lipsum}
\usepackage{booktabs,multirow}
\usepackage{placeins}
\usepackage{algpseudocode,algorithm}
\usepackage{color}
\usepackage[table,xcdraw]{xcolor}
\usepackage{bbm}
\usepackage{enumerate}   

\usepackage[normalem]{ulem}

\ninept
\def\reg{{\rm\ooalign{\hfil
      \raise.07ex\hbox{\scriptsize R}\hfil\crcr\mathhexbox20D}}}

\title{Multi-Frequency Information Enhanced Channel Attention Module for Speaker Representation Learning}

\makeatletter
\def\name#1{\gdef\@name{#1\\}}
\makeatother 
\name{{Mufan Sang, John H.L. Hansen} }

\address{
Center for Robust Speech Systems, University of Texas at Dallas, TX 75080}
\email{\small \tt \{mufan.sang, john.hansen\}@utdallas.edu}

\newcommand{\mf}[1]{{\color{blue}#1}}

\newcommand{\mfmod}[1]{{\color{red}{\emph{#1}}}}
\renewcommand{\mfmod}[1]{}

\graphicspath{{./image/}}

\begin{document}

\maketitle

\begin{abstract}
Recently, attention mechanisms have been applied successfully in neural network-based speaker verification systems. Incorporating the Squeeze-and-Excitation block into convolutional neural networks has achieved remarkable performance. However, it uses global average pooling (GAP) to simply average the features along time and frequency dimensions, which is incapable of preserving sufficient speaker information in the feature maps. In this study, we show that GAP is a special case of a discrete cosine transform (DCT) on time-frequency domain mathematically using only the lowest frequency component in frequency decomposition.  
\mfmod{Considering GAP as a special case of discrete cosine transform (DCT) with only using the lowest frequency component mathematically,}To strengthen the speaker information extraction ability, we propose to utilize multi-frequency information and design two novel and effective attention modules, called Single-Frequency Single-Channel (SFSC) attention module and Multi-Frequency Single-Channel (MFSC) attention module. The proposed attention modules can effectively capture more speaker information from multiple frequency components on the basis of DCT. We conduct comprehensive experiments on the VoxCeleb datasets and a probe evaluation on the $1^{\rm{st}}$48-UTD forensic corpus. Experimental results demonstrate that our proposed SFSC and MFSC attention modules can efficiently generate more discriminative speaker representations and outperform ResNet34-SE and ECAPA-TDNN systems with relative 20.9\% and 20.2\% reduction in EER, without adding extra network parameters.    
\end{abstract}



\vspace{-0.5ex}
\section{Introduction}
\label{sec:intro}
Speaker verification (SV) is a task aiming at identifying the true characteristics of a speaker and accepting or discarding the identity claimed by the speaker~\cite{hansen2015speaker}. In the past decade, we have seen the fast development of speaker recognition from the previous state-of-the-art i-vector system~\cite{dehak2010front} to deep neural network (DNN) based systems. \mfmod{Recently, great progresses have been made on DNN-based SV systems for supervised, semi-supervised, and self-supervised learning methods. }For supervised SV, various approaches 
were proposed with different deep neural network architectures~\cite{snyder2017deep, snyder2018x, desplanques2020ecapa}, novel loss functions~\cite{wan2018generalized, chung2020defence, wang2018cosface, deng2019arcface}, pooling methods~\cite{cai2018exploring, okabe2018attentive}, and frameworks for domain mismatch~\cite{ sang2020open, bhattacharya2019generative,sang2021deaan, wang2021multi}. In ~\cite{inoue2020semi, huh2020augmentation, sang2022self}, researchers further explored semi-supervised and self-supervised SV systems. 

Generally, an entire speaker verification system consists of a front-end speaker embedding network to extract discriminative speaker embeddings and a back-end scoring module to calculate the similarity or measure a score between embedding pairs. \mfmod{Therefore, improvements of either front-end speaker embedding network or back-end scoring part can lead to better speaker verification performance.}For speaker embedding networks, x-vector~\cite{snyder2018x} has proven its success and superiority over i-vector~\cite{dehak2010front} by utilizing the 1-D convolutional neural network (CNN) based time delay neural network (TDNN). Moreover, 2D CNNs (i.e. ResNet-based architectures) are also successfully applied to the SV task and produced remarkable performance~\cite{cai2018exploring, chung2020defence}. ECAPA-TDNN~\cite{desplanques2020ecapa} was proposed to further enhance the TDNN-based architecture and achieved a competitive performance with ResNet. \mfmod{These predominant CNN-based models take advantage of strong ability of extracting local temporal or frequency speaker patterns in speech.}To further improve the performance of networks, the attention mechanism is introduced to SV to enhance speaker information capture ability. The Squeeze-and-Excitation (SE) attention module~\cite{hu2018squeeze} was successfully adopted to speaker verification task and showed notable performance gain in~\cite{zhou2019deep}. \mfmod{It learns to weight the channels based on their importance.}The SE module collects global information of feature maps by using global average pooling (GAP) and it learns attentions to re-calibrate each channel by modeling the channel-wise relationships of features. \mfmod{In~\cite{yadav2020frequency}, researchers extended the Convolutional block attention module (CBAM)~\cite{woo2018cbam} from modelling channel-wise attention to both temporal and frequency attention independently.}Although SE-block has achieved superior performance on SV, simply averaging on temporal and frequency dimensions of feature maps may not sufficiently capture and preserve useful speaker information from input features. It could lead to the information loss problem and limit the representation power of CNNs. 

In this study, we propose a DCT-based multi-frequency information enhanced channel attention mechanism. Inspired by~\cite{qin2021fcanet}, we analyze the GAP pre-processing step on the time-frequency domain and show that GAP can be a special case of DCT-based frequency decomposition and it is actually equivalent to the lowest frequency component of DCT. Accordingly, we consider enhancing the speaker representation extraction ability by capturing comprehensive speaker information from multiple frequency components instead of only the lowest frequency component (i.e. GAP).\mfmod{utilizing multiple frequency components instead of only the lowest frequency component (i.e., GAP) to capture comprehensive speaker information from feature maps. }\mfmod{We aim to assign multiple frequency components instead of one single GAP value as indications of the  feature  channel  for  the  proceeding  importance  weight learning, which helped to produce diversified channel attention result.} Consequently, we propose two different DCT-based multi-frequency information enhanced channel attention modules: Single-Frequency Single-Channel attention (SFSC) and Multi-Frequency Single-Channel attention (MFSC). The SFSC attention module equally splits the input feature maps into several groups along the channel dimension to guarantee that each group can interact with one different DCT frequency component and produce one single importance indication from the corresponding frequency component. Furthermore, the MFSC attention module applies multiple different frequency components to the whole feature map without grouping and makes each channel fully interact with different frequency components of DCT. Experimental results on VoxCeleb datasets show that our proposed SFSC and MFSC attention modules can significantly outperform strong baselines ResNet34-SE and ECAPA-TDNN with relative 20.9\% and 20.2\% reduction in EER. Moreover, these advancements also show promise for naturalistic forensic speaker recognition. \mfmod{Moreover, they are still able to generate great performance in even more challenging scenarios for forensic speaker recognition.} 
\mfmod{The main contribution of our works are: (1) we proposed two efficient channel attention module which strengthen speaker information extraction ability by capturing information from diversified multiple frequency components. (2) we analyzed the connection between GAP and DCT-baed frequency decomposition. (3) We conduct comprehensive experiments to show that our attention modules can significantly outperform strong baseline models and they are still able to generate great performance in even more challenging forensic speaker recognition.  }




\mfmod{The rest of the paper is organized as follows. In Sec. 2, we introduce the frequency analysis and the details of our proposed attention modules. Data description, experimental setting, experimental results and discussion are reported in Sec. 3 and 4. Finally, we conclude this paper in Sec. 5.}

\begin{figure*}[th]
\centering
\scalebox{1.0}
{
\includegraphics[width=17.5cm,height=7.2cm]{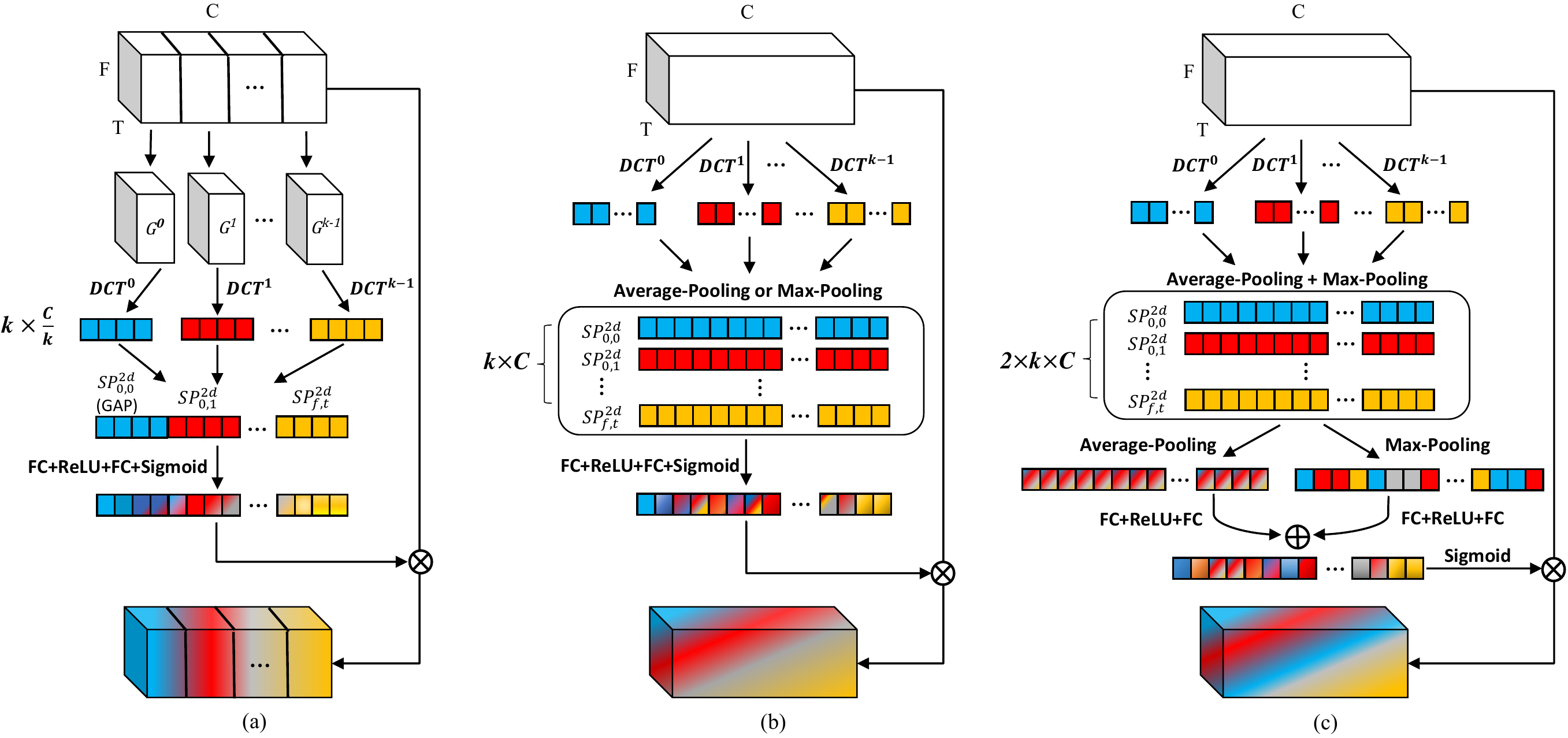}
}
\vspace{-6.0mm}
\caption{Overview of the proposed multi-frequency information enhanced channel attention modules: (a) The Single-Frequency Single-Channel (SFSC) attention module. (b) The Multi-Frequency Single-Channel (MFSC) attention module with average-pooling or max-pooling. (c) The Multi-Frequency Single-Channel (MFSC) attention module with average-pooling+max-pooling} 
\label{fig:system}
\end{figure*}

\vspace{-1.5ex}
\section{Methodology}
\label{sec:Method}
\subsection{Revisiting SE Attention}
\vspace{-1.0ex}
The channel attention mechanism has been successfully introduced to CNNs. Squeeze-and-excitation (SE) block~\cite{hu2018squeeze} models the interdependencies between the channels of feature maps with global information and recalibrate the feature maps to improve representation ability. It consists of squeeze and excitation two steps. For speech signals, the squeeze step applies GAP on temporal and frequency dimensions to generate channel-wise descriptors. Given $\mat{X}\in R^{C \times F \times T}$ as the feature map in networks, a channel-wise vector $\mathbf{Z}\in R^{C}$ is generated by GAP as:

\vspace{-1.5ex}
\begin{equation}
\begin{aligned}
{z}_{c}=GAP({x}_{c})=\frac{1}{F \times T} \sum_{i=1}^{F} \sum_{j=1}^{T} {x}_{c, i, j}
\end{aligned}
\end{equation}
where $C$, $F$, and $T$ represent the channel, frequency, and temporal dimensions. The scalar ${z}_{c}$ is the $c$-th element of $\mathbf{Z}$, denoting the mean statistics associated with the $c$-th channel. Then, the excitation step aims to capture channel-wise dependencies and output an attention vector by using two fully-connected layers $\mathbf{W}_{1}$ and $\mathbf{W}_{2}$ with a bottleneck architecture and non-linearity:

\vspace{-1ex}
\begin{equation}
\begin{aligned}
\mathbf{S}=\sigma\left(\mathbf{W}_{2} \delta\left(\mathbf{W}_{1} \mathbf{Z}\right)\right)
\end{aligned}
\end{equation}
where $\delta$ and $\sigma$ refer to ReLU and sigmoid activation function respectively. $\mathbf{S}\in R^{C}$ is the learned attention vector which dot multiplies to the original feature maps to re-scale each channel.

\vspace{-1ex}
\subsection{Thinking on Frequency Learning}
\mfmod{In SE-block, GAP aggregates the global information averages the time and frequency domain of feature maps to compress global information into a scalar for each channel/feature map. \mf{GAP aggregate the time and frequency domain information to capture global information into a scalar for each channel/feature map}}Although GAP has been commonly used in many attention mechanisms as a standard down-sampling method, we argue that simply utilizing average-pooling on temporal and frequency dimensions leads to insufficient information extraction from features and even information loss. To alleviate this problem, we conduct frequency component analysis and explore the connection between GAP and the result of frequency decomposition. 

Discrete cosine transform (DCT) is widely used in signal compression for both image and speech signals. It characterizes a sequence of data as a weighted sum of cosine functions oscillating at different
frequencies. For a raw speech signal $x\in R^{N}$ with length $N$, DCT can be formulated as:
\vspace{-1mm}
\begin{equation}
\begin{aligned}
SP_{k}=\sum_{i=0}^{N-1} x_{i} \cos \left(\frac{\pi k}{N}\left(i+\frac{1}{2}\right)\right), \text { s.t. } k \in\{0,1, \cdots, N-1\}\label{con:DCT1D}
\end{aligned}
\end{equation}
where $SP\in R^{N}$ is the frequency spectrum of DCT with $N$ different frequency components. For simplicity, we remove the constant coefficient of DCT. From Equation \ref{con:DCT1D}, we can find that the lowest frequency component ($k=0$) can be represented as:
\vspace{-1ex}
\begin{equation}
\begin{aligned}
SP_{0}=\sum_{i=0}^{N-1} x_{i}\label{con:DCT1D2}
\end{aligned}
\end{equation}
which is proportional to the result of GAP on the input data. We can extend Equation \ref{con:DCT1D} to the two-dimensional case for the signal $x\in R^{F \times T}$ in the time-frequency domain as:
\vspace{-1ex}
\begin{equation}
\begin{aligned}
SP^{2d}_{f,t}&=\sum_{i=0}^{F-1} \sum_{j=0}^{T-1} x_{i,j} \cos \left(\frac{\pi f}{F}\left(i+\frac{1}{2}\right)\right) \cos \left(\frac{\pi t}{T}\left(j+\frac{1}{2}\right)\right), \\
&\text { s.t. } f \in\{0,1, \cdots, F-1\}, t \in\{0,1, \cdots, T-1\} \label{con:DCT2D}
\end{aligned}
\end{equation}
in which $SP^{2d}_{f,t}\in R^{F \times T}$ denotes the 2D DCT frequency spectrum component at index $(f,t)$, $F$ and $T$ represent frequency and time dimensions of features. Derived from Equation \ref{con:DCT2D}, suppose $f$ and $t$ are both equal to 0, we have:
\begin{equation}
\begin{aligned}
SP^{2d}_{0,0} &=\sum_{i=0}^{F-1} \sum_{j=0}^{T-1} x_{i,j} \cos \left(\frac{\pi 0}{F}\left(i+\frac{1}{2}\right)\right) \cos \left(\frac{\pi 0}{T}\left(j+\frac{1}{2}\right)\right) \\
&= GAP(x_{i,j}) \times FT
\end{aligned}
\end{equation}

It indicates that the lowest frequency component $SP^{2d}_{0,0}$ of input features after 2D DCT is proportional to the result of GAP. As one of the decomposed frequency components, GAP may preserve only partial speaker information from the lowest frequency component among multiple frequency components.   
\mfmod{GAP is only one of the decomposed frequency components of input in frequency domain.}

On the other hand, input features can be decomposed into a weighted sum of multiple different frequency components by inverse DCT. We use $D^{i,j}_{f,t}$ to represent 2D DCT basis functions. The decomposed input features can be defined as below:
\begin{equation}
\begin{aligned}
D^{i,j}_{f,t}=\cos \left(\frac{\pi f}{F}\left(i+\frac{1}{2}\right)\right) \cos \left(\frac{\pi t}{T}\left(j+\frac{1}{2}\right)\right)
\end{aligned}
\end{equation}
\begin{equation}
\begin{aligned}
x_{i,j}&=\sum_{f=0}^{F-1} \sum_{t=0}^{T-1} SP^{2d}_{f,t} \cos \left(\frac{\pi f}{F}\left(i+\frac{1}{2}\right)\right) \cos \left(\frac{\pi t}{T}\left(j+\frac{1}{2}\right)\right)\\
&=SP^{2d}_{0,0}D^{i,j}_{0,0}+SP^{2d}_{0,1}D^{i,j}_{0,1}+\cdots+SP^{2d}_{F-1,T-1}D^{i,j}_{F-1,T-1}\\
&=GAP(x_{i,j}) F T+SP^{2d}_{0,1} D^{i,j}_{0,1}+\cdots+SP^{2d}_{F-1,T-1}D^{i,j}_{F-1,T-1}\label{con:inverse2D}
\end{aligned}
\end{equation}
\mfmod{we use $D^{i,j}_{f,t}$ to represent 2D DCT basis functions. }

As shown in Equation \ref{con:inverse2D}, speaker information of $x_{i,j}$ is distributed in different frequency components besides the lowest one. Therefore, the GAP pre-processing cannot extract adequate information from input features and lose much essential information contained in other frequency components.

\vspace{-1ex}
\subsection{Single-Frequency Single-Channel Attention Module}
Based on the analysis in Section 2.2, we aim to capture adequate speaker information from feature maps by exploiting more frequency components instead of only the lowest frequency. \mfmod{It is intuitively to expand GAP pre-processing to more frequency components.frequency components to pre-process features}Since the DCT weights are constant, it can be simply pre-calculated only once and saved in advance, which does not increase the training time and the number of network parameters. \mfmod{Therefore, we propose DCT-based multi-frequency attention module can be painlessly added to the existing deep neural network.}Therefore, we propose the single-frequency single-channel (SFSC) attention module which can be painlessly added to the existing speaker embedding networks. Figure \ref{fig:system} shows the overview of all our proposed attention modules. As shown in Figure \ref{fig:system}(a), SFSC equally splits the input feature maps along the channel dimension into several sub-groups as $[\mat{G}^{0},\mat{G}^{1},\cdots, \mat{G}^{k-1}]$ where $\mat{G}^{i}\in R^{\frac{C}{k} \times F \times T}$ and $k$ is the number of groups. Subsequently, each sub-group will be pre-processed by a corresponding DCT frequency component ranging from low frequency to high frequency. Every single channel within the same group is pre-processed by the same frequency component. Then, pre-processed sub-groups of input features are concatenated and the whole pre-processed feature map is represented as below:
\vspace{-1ex}
\begin{equation}
\begin{aligned}
z^{n}&=DCT_{f,t}(\mat{G}^{n})\\
&=\sum_{i=0}^{F-1} \sum_{j=0}^{T-1} D^{i,j}_{f,t}\mat{G}^{n}_{:,i,j}\\
\end{aligned}
\end{equation}
\begin{equation}
\begin{aligned}
\vspace{-2ex}
\mathbf{Z}_{SFSC}=cat([z^{0},z^{1},\cdots, z^{k-1}])\label{con:freq_cat_sfsc}
\end{aligned}
\end{equation}
in which $f$ and $t$ denotes the 2D indices of the frequency component corresponding to $\mat{G}^{n}$. And $z^{n} \in R^{\frac{C}{k}}$ is the feature vector after DCT pre-processing. \mfmod{Consequently, sub-groups of input features processed by multiple frequency components are concatenated and the whole input feature is represented as Equation \ref{con:freq_cat_sfsc}. }Once we obtain $\mathbf{Z}_{SFSC}$, the attention weights can be learned through two fully-connected layers with ReLU and sigmoid activation function as SE-Block. In this way, grouped features interact with different frequency components to acquire important speaker information comprehensively from frequency perspective. Furthermore, involving multiple frequency components encourages the network to enhance the diversity of extracted features.
\mfmod{Moreover, information redundancy exists in features of deep neural networks[ ]. It is possible to extract redundant speaker information if using GAP only. On the contrary, multiple frequency components may facilitate extracting complementary speaker information from redundant channels but within non-redundant frequency domain. frequency-encoded featureτcontains more diverse information pat-terns, which brings extra discriminative power to our model.This will be empirically discussed in }



\vspace{-1ex}
\subsection{Multi-Frequency Single-Channel Attention Module}
Effectively utilizing multiple frequency components to pre-process features is able to enrich preserved information for channel attention computing. For the proposed SFSC attention module in Section 2.3, every single channel of features is assigned with only one frequency component to produce one single channel indication. To further extend the interaction between channels and frequency components, we propose to assign multiple frequency components to every single channel without grouping. Figure \ref{fig:system}(b) and (c) show the mechanism of the proposed MFSC attention module. Different from SFSC, MFSC applies multiple different frequency components to the whole feature maps and makes each channel fully interact with multiple DCT frequency components. Given an entire feature map $\mat{X}\in R^{C \times F \times T}$, we have:
\begin{equation}
\begin{aligned}
\mathbf{Z}^{n}_{full}&=DCT_{f,t}(\mat{X})\\
&=\sum_{i=0}^{F-1} \sum_{j=0}^{T-1} D^{i,j}_{f,t}\mat{X}_{:,i,j}\\
\end{aligned}
\end{equation}
\vspace{-1ex}
\begin{equation}
\begin{aligned}
\mathbf{Z}_{MFSC}=M(cat([\mathbf{Z}^{0}_{full};\mathbf{Z}^{1}_{full};\cdots; \mathbf{Z}^{k-1}_{full}]))\label{con:freq_cat}
\end{aligned}
\end{equation}
where $\mathbf{Z}^{n}_{full}\in R^{C}$ denotes the feature vector of the entire feature map processed by the $n$-th DCT frequency component and we get totally $k$ feature vectors after pre-processing. The operation $M$ aggregates the stacked feature vectors processed by all $k$ frequency components and reduces the dimension from $k \times C$ to $C$. Here, we have three strategies for the operation $M$ applied to each channel across $k$ stacked vectors: (i) average-pooling, (ii) max-pooling, and (iii) average-pooling + max-pooling. We consider that max-pooling provides information from features in another aspect, and it is beneficial to finer channel-wise attention. For the third strategy, both aggregated vectors $\mathbf{Z}_{MFSC}$ from average-pooling and max-pooling are then forwarded to the following MLP in parallel to produce channel attention vectors. Before the sigmoid function, we merge these two output feature vectors by element-wise summation to obtain the channel attention finally as:
\vspace{-1ex}
\begin{equation}
\begin{aligned}
\mathbf{S}_{MFSC}=\sigma\left(\mathbf{W}_{2} \delta\left(\mathbf{W}_{1} \mathbf{Z}^{mean}_{MFSC}\right)+\mathbf{W}_{2} \delta\left(\mathbf{W}_{1} \mathbf{Z}^{max}_{MFSC}\right)\right)
\end{aligned}
\end{equation}
\mfmod{we apply $S$ as operating mean averaging or taking maximum to DCT frequency response for each channel across $k$ vectors. }

In this way, MFSC is able to embed diversified and complementary speaker information from multiple frequency components for each channel of features. 



\vspace{-1ex}
\section{Experimental Settings}
\subsection{Datasets and Feature Extraction}
All speaker verification systems are trained on the development set of VoxCeleb2~\cite{chung2018voxceleb2}, which consists of 1,092,009 utterances from 5994 speakers. We evaluate the performance of all systems on the VoxCeleb1~\cite{nagrani2017voxceleb} dataset, which contains over 100,000 utterances from 1251 speakers. We report the performance of systems on three evaluation trials: VoxCeleb1-O, VoxCeleb1-E, and VoxCeleb1-H. 

Moreover, we evaluate the proposed attention modules in more challenging scenarios on the naturalistic $1^{\rm{st}}$48-UTD forensic corpus~\cite{sang2020open}. \mfmod{The corpus is intended for forensic analysis of audio from various cities across the USA where detectives investigate actual homicides. Audio content was extracted from the USA TV program called “The First 48”, and all audio content are recorded from various real locations (e.g., interview rooms, cars, fields). }This corpus contains 300 speakers, and 5041 utterances consisting of 3.5 hours of actual situational crime audio. \mfmod{Each episode contains disjoint speakers, and speakers in every episode are tagged as Detective, Witness, and Suspect according to their identifies.}More than 50\% of utterances are shorter than 2 seconds. The test set contains 882 utterances from 39 speakers and the test trial has 7056 pairs. More details of the corpus can be found in~\cite{sang2020open}. 
\mfmod{It is a small domain-specific dataset with short utterances.  Audio consists of utterances with an average length of 2.4 s and more than 50\% of them are shorter than 2 seconds. Besides this duration constraint, context music, audio "bleeps" used for concealing harsh words, modified speech sound, and some voice-over are also contained in the audio. In this study, we use the training portion of the $1^{\rm{st}}$48-UTD corpus to fine-tune the trained student models obtained from Sec. 2.1, and evaluate its performance on the test portion. After filtering the utterances consisting of non-speech content and the speakers with fewer than three utterances, the training set consists of 3755 utterances from 228 speakers, and the test set contains 882 utterances from 39 speakers. }
\begin{table*}[t]
\vspace{-2mm}
\caption{The speaker verification performance of baseline systems, our proposed SFSC, MFSC and other SV systems on VoxCeleb1. avg: average-pooling, max: max-pooling}
\vspace{-2.5mm}
\setlength{\tabcolsep}{1.4mm}{
\renewcommand\arraystretch{1.2}
\scalebox{1.0}{
\begin{tabular}{cccccccccccc}
\hline \multirow{2}{*}{ Models } & \multirow{2}{*}{ Params (M) } & \multirow{2}{*}{ Loss function } & \multicolumn{2}{c}{ VoxCeleb1-O } & & \multicolumn{2}{c}{ VoxCeleb1-E } & & \multicolumn{2}{c}{ VoxCeleb1-H }\\
\cline { 4 - 5 } \cline { 7 - 8 } \cline { 10 - 11 }& & & EER(\%) & minDCF & & EER(\%) & minDCF & & EER(\%) & minDCF\\
\hline $ft$-CBAM {~\cite{yadav2020frequency}} & $4.6$ & AAM-softmax & $2.03$ & $-$ & & $-$ & $-$ & & $-$ & $-$\\ 
ECAPA-TDNN (C=512) & $6.2$ & AAM-softmax & $1.09$ & $0.084$ & & $1.30$ & $0.086$ & & $2.52$ & $0.156$\\
ResNet34-SE & $8.0$ & AP+softmax & $1.10$ & $0.079$ & & $1.28$ & $0.092$ & & $2.53$ & $0.162$\\
ResNet34-SE & $8.0$ & AAM-softmax & $1.16$ & $0.082$ & & $1.29$ & $0.087$ & & $2.41$ & $0.149$\\
\hline ResNet34-SFSC & $8.0$ & AP+softmax & $0.95$ & $0.073$ & & $1.19$ & $0.084$ & & $2.44$ & $0.158$\\
ResNet34-SFSC & $8.0$ & AAM-softmax & $0.97$ & $0.074$ & & $1.20$ & $0.081$ & & $\mathbf{2.29}$ & $\mathbf{0.142}$ \\
ResNet34-MFSC (avg) & $8.0$ & AP+softmax & $0.92$ & $0.072$ & & $1.13$ & $\mathbf{0.078}$ & & $2.31$ & $0.152$ \\
ResNet34-MFSC (avg) & $8.0$ & AAM-softmax & $0.95$ & $0.075$ & & $1.22$ & $0.081$ & & $2.33$ & $0.143$ \\
ResNet34-MFSC (max) & $8.0$ & AP+softmax & $0.94$ & $0.081$ & & $1.15$ & $0.080$ & & $2.31$ & $0.150$ \\
ResNet34-MFSC (max) & $8.0$ & AAM-softmax & $0.95$ & $0.076$ & & $1.24$ & $0.082$ & & $2.35$ & $0.150$ \\
ResNet34-MFSC (avg+max) & $8.0$ & AP+softmax & $\mathbf{0.87}$ & $\mathbf{0.068}$ & & $\mathbf{1.11}$ & $0.079$ & & $2.32$ & $0.148$ \\
ResNet34-MFSC (avg+max) & $8.0$ & AAM-softmax & $0.91$ & $0.071$ & & $1.21$ & $0.083$ & & $2.30$ & $0.146$\\
\hline
\end{tabular}}}
\label{tab:vox1perf}
\end{table*}

For all the systems, we compute 64-dimensional log Mel-filterbanks with a frame-length of 25 ms and 10 ms shift. We apply mean and variance normalization (MVN) to the input. \mfmod{VAD is not applied since training data in the VoxCeleb mostly consists of continuous speech. }For each input utterance, a segment of 2.0 seconds is randomly selected.

\vspace{-1ex}
\subsection{Data Augmentation}
\vspace{-1ex}
Online data augmentation has shown its successes in supervised and self-supervised speaker verification~\cite{cai2020fly, sang2022self}. \mfmod{The increasing amount and variability of training data can improve the robustness of a speaker embedding system.}Different from conventional offline data augmentation strategies used in SV, our online augmentation allows multiple different augmentations randomly and collectively applied to each input utterance. We apply additive noise with MUSAN~\cite{snyder2015musan}, reverberation with RIR~\cite{ko2017study}, and SpecAug~\cite{park2019specaugment} collectively for augmentation. We follow the setting in \cite{sang2022self} for the online data augmentation.  

\vspace{-3ex}
\subsection{Model Training and Implementation Details}
\vspace{-1ex}
We exploit the ResNet34 in~\cite{heo2020clova} with attentive statistic pooling (ASP)~\cite{okabe2018attentive} as the backbone network. Our re-implemented ResNet34-SE and ECAPA-TDNN are used as baselines. In this study, we use 16 DCT frequency components ($k$=16) based on the size of feature maps from the last residual block of ResNet34 for all SFSC and MFSC modules. 

For model training, we use the Adam optimizer~\cite{kingma2014adam} with an initial learning rate of 1e-3. The learning rate is reduced to 1e-5 with a 0.75 decay in every 15 epochs. We use two different loss functions to train all the models: (i) the additive angular margin softmax (AAM-softmax)~\cite{deng2019arcface} with a margin $m$ of 0.2 and a scale factor $s$ of 30, (ii) and the combined angular prototypical loss together with the softmax loss~\cite{chung2020defence} (AP+softmax). The size of extracted speaker embeddings is 512. During training, the batch size of 200 is used for the AAM-softmax loss and 280 for the AP+softmax loss. 

For evaluation, we report the system performances using two evaluation metrics: Equal Error Rate (EER) and minimum Detection Cost Function (minDCF) with $p_{target}$= 0.05. Cosine similarity is adopted for scoring in the testing phase.

\vspace{-1.0ex}
\section{Experimental Results and Discussions}
\vspace{-1.0ex}
Table \ref{tab:vox1perf} presents the performance of all the systems on VoxCeleb1 dataset. Compared to the ResNet34-SE and ECAPA-TDNN, our proposed SFSC and MFSC attention modules significantly outperform them in both EER and minDCF on VoxCeleb1-O, VoxCeleb1-E, and VoxCeleb1-H. From Table 1, we observe that the ResNet34-SFSC trained by AP+softmax and AAM-softmax losses can improve the performance with relative 13.6\% and 16.4\% reduction in EER compared to ResNet34-SE on VoxCeleb1-O. The proposed MFSC with ResNet34 trained by AP+softmax and AAM-softmax can further boost the performance to 0.87\% and 0.91\% EER on VoxCeleb1-O, with relative 20.9\% and 21.6\% reduction in EER. For MFSC, Table \ref{tab:vox1perf} shows the performance of models trained with three different aggregation strategies: average-pooling, max-pooling, and average-pooling + max-pooling. We observe that MFSC with average-pooling can generate slightly better performance than using max-pooling. And MFSC with average-pooling+max-pooling aggregation achieves the best performance. Our experimental results indicate that utilizing more frequency components instead of only the lowest one (i.e. GAP) encourages CNN-based models to extract sufficient speaker information from feature maps. Moreover, it is worth noting that our SFSC and MFSC attention modules provide remarkable performance gain to the whole speaker embedding network without adding any extra network parameters.  

\vspace{-2mm}
\begin{table}[h]
\caption{The performance of ResNet34-SE, ResNet34-SFSC and ResNet34-MFSC models on the $1^{st}$48-UTD forensic dataset.}
\vspace{-6mm}
\setlength{\tabcolsep}{1.2mm}{
\begin{center}
\renewcommand\arraystretch{1.1}
\scalebox{1.0}{
\begin{tabular}{ccc}
\hline \textbf{Models} & \textbf{EER(\%)} & \textbf{minDCF} \\
\hline ResNet34-SE & $16.31$ & $0.725$ \\
\hline ResNet34-SFSC & $15.34$ & $0.650$ \\
ResNet34-MFSC & $\mathbf{14.97}$ & $\mathbf{0.643}$ \\
\hline
\end{tabular}}
\end{center}}
\vspace{-3mm}
\label{tab:1st48perf}
\end{table}

\vspace{-3mm}
Moreover, we evaluate the performance of proposed attention modules in more challenging scenarios for forensic speaker recognition, where audio contents are recorded from various real locations (e.g., police interview rooms, cars, fields). Table \ref{tab:1st48perf} shows the results of ResNet34-SE and our attention models evaluated on the $1^{\rm{st}}$48-UTD dataset. Compared to ResNet34-SE, we observe that the SFSC and MFSC decrease the EER from 16.31\% to 15.34\% and 14.97\% and minDCF from 0.725 to 0.650 and 0.643, respectively. It suggests that our attention modules are more speaker-discriminative and robust even in more complex scenarios for speaker verification.
\mfmod{To visualize the effectiveness of the proposed model in embedding space, we use the t-distributed Stochastic Neighbor Embedding (t-SNE) plots to illustrate the performance improvement in Figure 2. Ten speakers are randomly selected from VoxCeleb1 test set. \mfmod{Figure 2(a),(b),(c) visualize the speaker embeddings extracted by ResNet34-SE, ResNet34-SFSC, and ResNet34-MFSC, respectively.} As shown in Figure 2, it can be observed that speaker embeddings extracted from ResNet34-SFSC and ResNet34-MFSC are denser for same speaker and more separate for different speakers, which demonstrates the proposed attention modules is beneficial to speaker embedding networks to extracting more discriminative speaker embeddings by collaborating features from multiple frequency components. }

\section{Conclusion}
In this paper, we propose multi-frequency information enhanced channel attention modules: single-frequency single-channel (SFSC) attention and multi-frequency single-channel (MFSC) attention. Without adding extra network parameters, both of them show the effectiveness of incorporating multiple frequency components to facilitate the representation extraction ability for speaker embedding networks. SFSC captures information from multiple frequency components with feature grouping and each group interacts with one single frequency component. MFSC incorporates comprehensive multi-frequency information by fully interacting with multiple frequency components. For MFSC, using average-pooling and max-pooling collectively contributes to a more efficient attention module with a stronger representation ability. Experimental results demonstrate that the proposed attention modules can significantly outperform all strong baselines and they are still efficient for naturalistic forensic speaker recognition.

\vfill\pagebreak

\newpage

\bibliographystyle{IEEEtran}
\bibliography{is2019_sed.bib}

\end{document}